\documentclass[12pt]{article}
\usepackage{epsf}
\def\beq{\begin{equation}}
\def\eeq{\end{equation}}

\def\ap#1#2#3 {Ann. Phys. (NY) {\bf#1} (19#2) #3}
\def\err#1#2#3 {{\it Erratum} {\bf#1} (19#2) #3}
\def\ib#1#2#3 {{\it ibid.} {\bf#1} (19#2) #3}
\def\ijmp#1#2#3 {Int. J. Mod. Phys. {\bf#1} (19#2) #3}
\def\jetp#1#2#3 {JETP Lett. {\bf#1} (19#2) #3}
\def\mpl#1#2#3 {Mod. Phys. Lett. {\bf#1} (19#2) #3}
\def\np#1#2#3 {Nucl. Phys. {\bf#1} (19#2) #3}
\def\pl#1#2#3 {Phys. Lett. {\bf#1} (19#2) #3}
\def\prep#1#2#3 {Phys. Rep. {\bf#1} (19#2) #3}
\def\prev#1#2#3 {Phys. Rev. {\bf#1} (19#2) #3}
\def\prl#1#2#3 {Phys. Rev. Lett. {\bf#1} (19#2) #3}
\def\sjnp#1#2#3 {Sov. J. Nucl. Phys. {\bf#1} (19#2) #3}
\def\spj#1#2#3 {Sov. Phys. JETP {\bf#1} (19#2) #3}
\def\spu#1#2#3 {Sov. Phys. Usp. {\bf#1} (19#2) #3}
\def\zp#1#2#3 {Zeit. Phys. {\bf#1} (19#2) #3}

\begin{document}
\begin{titlepage}
\begin{center}
{\Large \bf Theoretical Physics Institute \\
University of Minnesota \\}
\end{center}
\vspace{0.2in}
\begin{flushright}
TPI-MINN-00/17 \\
UMN-TH-1851-00 \\
%hep-ph/0004257\\
April 2000 \\
\end{flushright}
\vspace{0.3in}
\begin{center}
{\Large \bf Inclusive weak decay rates of heavy hadrons
\\}
\vspace{0.2in}
{\bf M.B. Voloshin  \\ }
Theoretical Physics Institute, University of Minnesota, Minneapolis,
MN
55455 \\ and \\
Institute of Theoretical and Experimental Physics, Moscow, 117259
\\[0.2in]
{\it Expanded version of a contribution to the final report book of the
Fermilab Workshop on B Physics at Tevatron}
\end{center}

\begin{abstract}

A compact review of the theory, including some recent developments, of
inclusive weak decay rates of charmed and $b$ hadrons with an emphasis
on predictions that can be tested in the forthcoming experiments.

\end{abstract}

%\noindent
%PACS: 12.39.Hg, 14.20.Lq, 14.20.Mr, 13.30.-a

\end{titlepage}

\section{Introduction}
The dominant weak decays of hadrons containing a heavy quark, $c$ or
$b$, are caused by the decay of the heavy quark. In the limit of a very
large mass $m_Q$ of a heavy quark $Q$ the parton picture of the hadron
decay should set in, where the inclusive decay rates of hadrons,
containing $Q$, mesons ($Q\bar q$) and baryons ($Qqq$), are all the same
and equal to the inclusive decay rate $\Gamma_{parton}(Q)$ of the heavy
quark. Yet, the known inclusive decay rates \cite{pdg} are conspicuously
different for different hadrons, especially for charmed hadrons, whose
lifetimes span a range of more than one order of magnitude from the
shortest $\tau(\Omega_c)=0.064 \pm 0.020$ ps to the longest
$\tau(D^+)=1.057 \pm 0.015$ ps, while the differences of lifetime among
$b$ hadrons are substantially smaller. The relation between the relative
lifetime differences for charmed and $b$ hadrons reflects the fact that
the dependence of the inclusive decay rates on the light quark-gluon
`environment' in a particular hadron is a pre-asymptotic effect in the
parameter $m_Q$, which effect vanishes as an inverse power of $m_Q$ at
large mass.

A theoretical framework for systematic description of the leading at
$m_Q \to \infty$ term in the inclusive decay rate $\Gamma_{parton}(Q)
\propto m_Q^5$ as well as of the terms relatively suppressed by inverse
powers of $m_Q$ is provided \cite{sv:81,sv:85,sv:86} by the operator
product expansion (OPE) in $m_Q^{-1}$. Existing theoretical predictions
for inclusive weak decay rates are in a reasonable agreement, within the
expected range of uncertainty, with the data on lifetimes of charmed
particles and with the so far available data on decays of $B$ mesons.
The only outstanding piece of present experimental data is on the
lifetime of the $\Lambda_b$ baryon: $\tau(\Lambda_b)/\tau(B_d) \approx
0.8$, for which ratio a theoretical prediction, given all the
uncertainty involved, is unlikely to produce a number lower than 0.9.
The number of available predictions for inclusive decay rates of charmed
and $b$ hadrons is sufficiently large for future experimental studies to
firmly establish the validity status of the OPE based theory of heavy
hadron decays, and, in particular, to find out whether the present
contradiction between the theory and the data on
$\tau(\Lambda_b)/\tau(B_d)$ is a temporary difficulty, or an evidence of
fundamental flaws in theoretical understanding.

It is a matter of common knowledge that application of  OPE to decays of
charmed and $b$ hadrons has potentially two caveats. One is that the OPE
is used in the Minkowsky kinematical domain, and therefore relies on the
assumption
of quark-hadron duality at the energies involved in the corresponding
decays. In other words, it is assumed that sufficiently many exclusive
hadronic channels contribute to the inclusive rate, so that the
accidentals of the low-energy resonance structure do not affect the
total rates of the inclusive processes. Theoretical attempts at
understanding the onset of the quark-hadron duality are so far limited
to model estimates \cite{cdsu:97,bsuv:99}, not yet suitable for direct
quantitative evaluation of possible deviation from duality in charm and
$b$ decays. This point presents the most fundamental uncertainty of the
OPE based approach, and presently can only be clarified by confronting
theoretical predictions with experimental data. The second possible
caveat in applying the OPE technique to inclusive charm decays is that
the mass of the charm quark, $m_c$, may be insufficiently large for
significant suppression of higher terms of the expansion in $m_c^{-1}$.
The relative lightness of the charm quark, however, accounts for a
qualitative, and even semi-quantitative, agreement of the OPE based
predictions with the observed large spread of the lifetimes of charmed
hadrons: the nonperturbative effects, formally suppressed by $m_c^{-2}$
and $m_c^{-3}$ are comparable with the `leading' parton term and
describe the hierarchy of the lifetimes.

Another uncertainty of a technical nature arises from poor knowledge of
matrix elements of certain quark operators over hadron, arising as terms
in OPE. These can be estimated within theoretical models, with
inevitable ensuing model dependence, or, where possible, extracted from
the experimental data.
With these reservations spelled out, we discuss here the OPE based
description of inclusive weak decays of charm and $b$ hadrons, with
emphasis on specific experimentally testable predictions, and on the
measurements, which would less rely on model dependence of the estimates
of
the matrix elements, thus allowing to probe the OPE predictions at a
fundamental level.

\section{OPE for inclusive weak decay rates}
The optical theorem of the scattering theory relates the total decay
rate $\Gamma_H$ of a hadron $H_Q$ containing a heavy quark $Q$ to the
imaginary part of the `forward scattering amplitude'. For the case of
weak decays the latter amplitude is described by the following effective
operator
\beq
L_{eff}=2 \,{\rm Im} \, \left [ i \int d^4x \, e^{iqx} \, T \left \{
L_W(x),
L_W(0) \right \} \right ]~,
\label{leff}
\eeq
in terms of which the total decay rate is given by\footnote{We use here
the non-relativistic normalization
for the {\it heavy} quark states: $\langle
Q | Q^\dagger Q | Q \rangle =1$.}
\beq
\Gamma_H=\langle H_Q | \, L_{eff} \, | H_Q \rangle~.
\label{lgam}
\eeq
The correlator in equation (\ref{leff}) in general is a non-local
operator. However at $q^2=m_Q^2$ the dominating space-time intervals in
the integral are of order $m_Q^{-1}$ and one can expand the correlator
in $x$, thus producing an expansion in inverse powers of $m_Q$. The
leading term in this expansion describes the parton decay rate of the
quark. For instance, the term  in the non-leptonic weak Lagrangian
$\sqrt{2} \, G_F
\, V ({\overline q}_{1 L} \gamma_\mu \, Q_L)({\overline q}_{2 L}
\gamma_\mu \, q_{3 L})$ with $V$ being the appropriate combination of
the CKM mixing factors, generates through eq.(\ref{leff}) the leading
term in the effective Lagrangian
\beq
L^{(0)}_{eff, \, nl} = |V|^2 \, {G_F^2 \, m_Q^5 \over 64 \, \pi^3} \,
\eta_{nl} \, \left ( {\overline Q} Q \right )~,
\label{lef0}
\eeq
where $\eta_{nl}$ is the perturbative QCD radiative correction factor.
This expression reproduces the well known formula for the inclusive
non-leptonic decay rate of a heavy quark, associated with the underlying
process $Q \to q_1 \, q_2 \, {\overline q}_3$, due to the relation
$\langle H_Q
| {\overline Q} Q | H_Q \rangle \approx \langle H_Q | Q^\dagger Q | H_Q
\rangle =1$, which is valid up to corrections of order $m_Q^{-2}$. One
also sees form this example, that in order to separate individual
semi-inclusive decay channels, e.g. non-leptonic with specific flavor
quantum numbers, or semi-leptonic, one should simply pick up the
corresponding relevant part of the weak Lagrangian $L_W$, describing the
underlying process, to include in the correlator (\ref{leff}).

The general expression for first three terms in the OPE for $L_{eff}$
has the form
\begin{eqnarray}
&&L_{eff}=
L_{eff}^{(0)} + L_{eff}^{(2)} + L_{eff}^{(3)}= \nonumber \\
&& c^{(0)} \, {G_F^2 \, m_Q^5 \over 64 \, \pi^3} \, \left ( {\overline
Q} Q \right ) +  c^{(2)} \,{G_F^2 \, m_Q^3 \over 64 \, \pi^3} \, \left
({\overline Q}
\, \sigma^{\mu \nu} G_{\mu \nu} \, Q \right ) + {G_F^2 \, m_Q^2 \over 4
\, \pi} \,\sum_i c_i^{(3)} \,
({\overline q}_i \Gamma_i
q_i)({\overline Q} \Gamma^\prime_i Q)~,
\label{first3}
\end{eqnarray}
where the superscripts denote the power of $m_Q^{-1}$ in the relative
suppression of the corresponding term in the expansion  with respect to
the leading one, $G_{\mu \nu}$ is the gluon field tensor, $q_i$ stand
for light quarks, $u,\, d, \, s$, and, finally, $\Gamma_i$,
$\Gamma^\prime_i$ denote spin and color structures of the four-quark
operators. The coefficients $c^{(a)}$ depend on the specific part of the
weak interaction Lagrangian $L_W$, describing the relevant underlying
quark process.

One can notice the absence in the expansion (\ref{first3}) of a term
suppressed by just one power of $m_Q^{-1}$, due to non-existence of
operators of suitable dimension. Thus the decay rates receive no
correction of relative order $m_Q^{-1}$ in the limit of large $m_Q$, and
the first pre-asymptotic corrections appear only in the order
$m_Q^{-2}$.

The mechanisms giving rise to the three discussed terms in OPE are shown
in Figure 1. The first, leading term corresponds to the
parton decay, and does not depend on the light guark and gluon
`environment'  of the heavy quark in a hadron. The second term describes
the effect on the decay rate of the gluon field that a heavy quark
`sees' in a hadron. This term in fact is sensitive only to the
chromomagnetic part of the gluon field, and contains the operator of the
interaction of heavy quark chromomagnetic moment with the chromomagnetic
field. Thus this term depends on the spin of the heavy quark, but does
not depend on the flavors of the light quarks or antiquarks. Therefore
this effect does not split the inclusive decay rates within flavor SU(3)
multiplets of heavy hadrons, but generally gives difference of the
rates, say, between mesons and baryons. The dependence on the light
quark flavor arises from the third term in the expansion (\ref{first3})
which explicitly contains light quark fields. Historically, this part is
interpreted in terms of two mechanisms \cite{sv:81,gnpr:79,bgt:84}: the
weak scattering (WS) and the Pauli interference (PI). The WS corresponds
to a cross-channel of the underlying decay, generically $Q \to q_1 \,
q_2 \, {\overline
q}_3$, where either the quark $q_3$ is a spectator in a baryon and can
undergo a weak scattering off the heavy quark: $q_3 \, Q \to q_1 \,
q_2$, or an antiquark in meson, say ${\overline q}_1$, weak-scatters
(annihilates) in the process ${\overline q}_1 \, Q \to q_2 \, {\overline
q}_3$. The Pauli interference effect arises when one of the final
(anti)quarks in the decay of $Q$ is identical to the spectator
(anti)quark in the hadron, so that an interference of identical
particles should be taken into account. The latter interference can be
either constructive or destructive, depending on the relative spin-color
arrangement of the (anti)quark produced in the decay and of the
spectator one, thus the sign of the PI effect is found only as a result
of specific dynamical calculation. In specific calculations, however, WS
and PI arise from the same terms in OPE, depending on the hadron
discussed, and technically there is no need to resort to the traditional
terminology of WS and PI.

\begin{figure}[ht]
\begin{center}
\unitlength 0.9mm
\thicklines
\begin{picture}(140.00,84.00)(10,0)
\put(10.00,70.00){\line(1,0){20.00}}
\put(30.00,70.00){\line(1,0){30.00}}
\put(60.00,70.00){\line(1,0){20.00}}
\bezier{164}(30.00,70.00)(45.00,84.00)(60.00,70.00)
\bezier{164}(30.00,70.00)(45.00,56.00)(60.00,70.00)
\put(30.00,70.00){\circle*{2.00}}
\put(60.00,70.00){\circle*{2.00}}
\put(20.00,72.00){\makebox(0,0)[cb]{$Q$}}
\put(70.00,71.00){\makebox(0,0)[cb]{$Q$}}
\put(49.00,78.00){\makebox(0,0)[cb]{$q_1$}}
\put(49.00,71.00){\makebox(0,0)[cb]{$q_2$}}
\put(49.00,62.00){\makebox(0,0)[ct]{${\overline q_3}$}}
\put(10.00,43.00){\line(1,0){20.00}}
\put(30.00,43.00){\line(1,0){30.00}}
\put(60.00,43.00){\line(1,0){20.00}}
\bezier{164}(30.00,43.00)(45.00,57.00)(60.00,43.00)
\bezier{164}(30.00,43.00)(45.00,29.00)(60.00,43.00)
\put(30.00,43.00){\circle*{2.00}}
\put(60.00,43.00){\circle*{2.00}}
\put(20.00,45.00){\makebox(0,0)[cb]{$Q$}}
\put(70.00,44.00){\makebox(0,0)[cb]{$Q$}}
\put(49.00,51.00){\makebox(0,0)[cb]{$q_1$}}
\put(49.00,44.00){\makebox(0,0)[cb]{$q_2$}}
\put(49.00,35.00){\makebox(0,0)[ct]{${\overline q_3}$}}
\put(45.00,36.00){\line(0,-1){2.00}}
\put(45.00,33.00){\line(0,-1){2.00}}
\put(45.00,30.00){\line(0,-1){2.00}}
\put(47.00,28.00){\makebox(0,0)[cc]{$g$}}
\put(10.00,12.00){\line(1,0){20.00}}
\put(30.00,12.00){\line(1,0){30.00}}
\put(60.00,12.00){\line(1,0){20.00}}
\bezier{164}(30.00,12.00)(45.00,26.00)(60.00,12.00)
\put(30.00,12.00){\circle*{2.00}}
\put(60.00,12.00){\circle*{2.00}}
\put(20.00,14.00){\makebox(0,0)[cb]{$Q$}}
\put(70.00,13.00){\makebox(0,0)[cb]{$Q$}}
\put(49.00,20.00){\makebox(0,0)[cb]{$q_1$}}
\put(49.00,13.00){\makebox(0,0)[cb]{$q_2$}}
\put(10.00,5.50){\line(3,1){20.00}}
\put(60.00,12.00){\line(4,-1){20.00}}
\put(20.00,7.00){\makebox(0,0)[ct]{$q_3$}}
\put(70.00,8.00){\makebox(0,0)[ct]{$q_3$}}
\put(87.00,70.00){\vector(1,0){11.00}}
\put(87.00,43.00){\vector(1,0){11.00}}
\put(87.00,12.00){\vector(1,0){11.00}}
\put(110.00,70.00){\line(1,0){30.00}}
\put(116.00,71.00){\makebox(0,0)[cb]{$Q$}}
\put(134.00,71.00){\makebox(0,0)[cb]{$Q$}}
\put(110.00,43.00){\line(1,0){30.00}}
\put(116.00,44.00){\makebox(0,0)[cb]{$Q$}}
\put(134.00,44.00){\makebox(0,0)[cb]{$Q$}}
\put(125.00,43.00){\line(0,-1){2.00}}
\put(125.00,40.00){\line(0,-1){2.00}}
\put(125.00,37.00){\line(0,-1){2.00}}
\put(127.00,35.00){\makebox(0,0)[cc]{$g$}}
\put(110.00,12.00){\line(1,0){30.00}}
\put(116.00,13.00){\makebox(0,0)[cb]{$Q$}}
\put(134.00,13.00){\makebox(0,0)[cb]{$Q$}}
\put(110.00,7.00){\line(3,1){15.00}}
\put(125.00,12.00){\line(3,-1){15.00}}
\put(117.00,9.00){\makebox(0,0)[lt]{$q$}}
\put(132.00,9.00){\makebox(0,0)[rt]{$q$}}
\put(125.00,62.00){\makebox(0,0)[cc]{ $m_Q^5 \, ({\overline Q} Q)$}}
\put(125.00,27.00){\makebox(0,0)[cc]{ $ m_Q^3 \, \left ({\overline Q}
(\vec \sigma \cdot \vec B) Q \right )$}}
\put(125.00,0.00){\makebox(0,0)[cc]{ $m_Q^2 \, ({\overline q} \Gamma
q)({\overline Q} \Gamma^\prime Q)$}}
\put(123.50,70.00){ \circle*{2.83}}
\put(123.50,43.00){ \circle*{2.83}}
\put(123.50,12.00){ \circle*{2.83}}
\end{picture}
\caption{Graphs for three first terms in OPE for inclusive decay rates:
the parton term, the chromomagnetic interaction, and the four-quark
term.}
\end{center}
\label{fig:ope}
\end{figure}
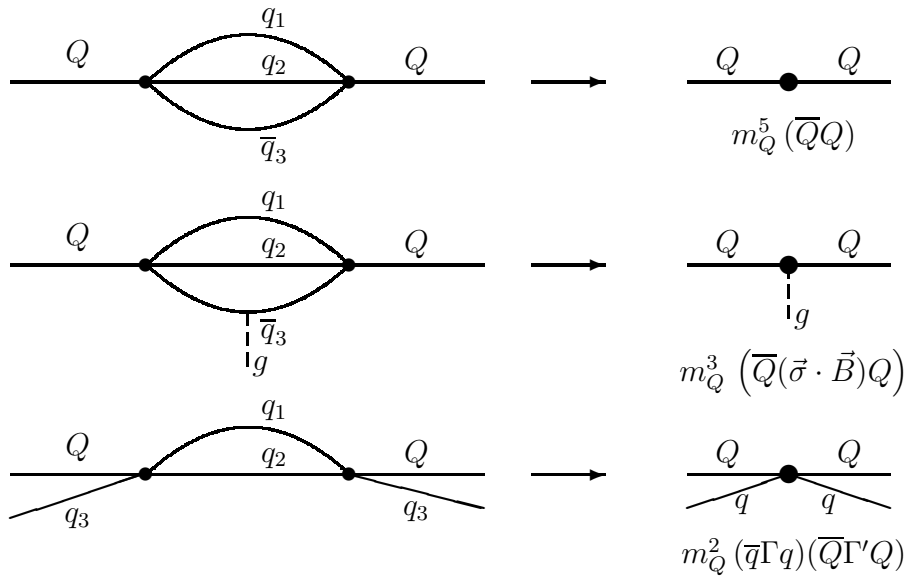

In what follows we discuss separately the effects of the three terms in
the expansion (\ref{first3}) and their interpretation within the
existing and future data.

\section{The parton decay rate}
The leading term in the OPE amounts to the perturbative expression for
the decay rate of a heavy quark. In $b$ hadrons the contribution of the
subsequent terms in OPE is at the level of few percent, so that the
perturbative part can be confronted with the data in its own right. In
particular, for the $B_d$ meson the higher terms in OPE contribute only
about 1\%
of the total non-leptonic as well as of the semileptonic decay rate.
Thus the data on these rates can be directly compared with the leading
perturbative term in OPE.

The principal theoretical topic, associated with this term is the
calculation of QCD radiative corrections, i.e. of the factor $\eta_{nl}$
in eq.(\ref{lef0}) and of a similar factor, $\eta_{nl}$, for
semileptonic decays. It should be noted, that even at this,
perturbative, level there is a known long-standing problem between the
existing data and the theory in that the current world average for the
semileptonic branching ratio for the $B$ mesons, $B_{sl}(B)= 10.45 \pm
0.21 \%$, is somewhat lower than the value $B_{sl}(B) \ge 11.5$
preferred from the present knowledge of theoretical QCD radiative
corrections to the ratio of non-leptonic to semileptonic decay rates
(see e.g. \cite{bbsv:94}). However, this apparent discrepancy may in
fact be due to insufficient `depth' of perturbative QCD calculation of
the ratio $\eta_{nl}/\eta_{sl}$. In order to briefly elaborate on this
point, we notice that the standard way of analyzing the perturbative
radiative corrections in the nonleptonic decays is through the
renormalization group (RG) summation of the leading log terms and the
first next-to-leading terms \cite{ap:91,bbbg:95} in the parameter $L
\equiv \ln(m_W/m_b)$. For the semileptonic decays the logarithmic
dependence on $m_W/m_b$ is absent in all orders due to the weak current
conservation at momenta larger than $m_b$, thus the correction is
calculated by the standard perturbative technique, and a complete
expression in the first order in $\alpha_s$ is available both for the
total rate \cite{hp:83,nir:89} and for the lepton spectrum \cite{cj:94}.
In reality however the parameter $L \approx 2.8$ is not large, and
non-logarithmic terms may well compete with the logarithmic ones. This
behavior is already seen from the known expression for the logarithmic
terms: when expanded up to the order $\alpha_s^2$ the result of
Ref.\cite{bbbg:94} for the rate of decays with single final charmed
quark takes the form
\beq
{{\Gamma(b \to c \bar u d) +
\Gamma(b \to c \bar u s)} \over {3 \, \Gamma(b \to c e
\bar \nu )}}=
 1+ {\alpha_s \over \pi} + {\alpha_s^2 \over \pi^2} \, \left [ 4 \, L^2
+ \left
( {7 \over 6} + { 2 \over 3} \, c(m_c^2/m_b^2) \right ) \, L \right ]~,
\label{as2l}
\eeq
where, in terms of notation of Ref.\cite{bbbg:94},
$c(a)=c_{22}(a)-c_{12}(a)$. The behavior of the function $c(a)$ is known
explicitly \cite{bbbg:94} and is quite weak:  $c(0)=19/2$, $c(1) = 6$,
and $c(m_c^2/m_b^2) \approx 9.0$ for the realistic mass ratio $m_c/m_b
\approx 0.3$. One can see that the term with the single logarithm $L$
contributes about two thirds of that with $L^2$ in the term quadratic in
$\alpha_s$. Under such circumstances the RG summation of the terms with
powers of $L$ does not look satisfactory for numerical estimates of the
QCD effects, at least at the so far considered level of the first
next-to-leading order terms, and the next-to-next-to-leading terms can
be equally important as the two known ones, which would eliminate the
existing impasse between the theory and the data on $B_{sl}(B)$. One can
present some arguments \cite{mv:96} that this is indeed the case for the
$b$ quark decay, although a complete calculation of these corrections is
still unavailable.

\section{Chromomagnetic and time dilation effects in decay rates}

The corrections suppressed by two powers of $m_Q^{-1}$ to inclusive
decay rates arise from two sources \cite{buv:92}: the $O(m_Q^{-2})$
corrections to the matrix element of the leading operator, $({\overline
Q} Q$, and the second term in OPE (\ref{first3}) containing the
chromomagnetic interaction. The expression for the matrix element of the
leading operator with the correction included is written in the form
\beq
\langle H_Q| {\overline Q} Q | H_Q \rangle = 1-
{\mu_\pi^2(H_Q)-\mu_g^2(H_Q) \over 2 \, m_Q^2} + \ldots ,
\label{qbarq}
\eeq
where $\mu_\pi^2$ and $\mu_g^2$ are defined as
\begin{eqnarray}
&&\mu_\pi^2=\langle H_Q| {\overline Q}\, (i {\vec D})^2 \, Q | H_Q
\rangle ~, \nonumber \\
&&\mu_g^2=\langle H_Q| {\overline Q}\, {1 \over 2} \sigma^{\mu \nu}
G_{\mu \nu} Q | H_Q \rangle~,
\label{mus}
\end{eqnarray}
with $D$ being the QCD covariant derivative. The correction in equation
(\ref{qbarq}) in fact corresponds to the time dilation factor $m_Q/E_Q$,
for the heavy quark decaying inside a hadron, where it has energy $E_Q$,
which energy is contributed by the kinetic part ($\propto \mu_\pi^2$)
and the chromomagnetic part ($\propto \mu_g^2$). The second term in OPE
describes the effect of the chromomagnetic interaction in the decay
process, and is also expressed through $\mu_g^2$.

The explicit formulas for the decay rates, including the effects up to
the order $m_Q^{-2}$ are found in \cite{buv:92} and for decays of the
$b$ hadrons read as follows. For the semileptonic decay rate
\beq
\Gamma_{sl}(H_b)={|V_{cb}|^2 \, G_F^2 \, m_b^5 \over 192 \, \pi^3}
\,\langle H_b | {\overline b} b | H_b \rangle \,\left [1+
 {{\mu_g^2} \over m_b^2} \left ( {x \over 2}
\,{ d \over
{dx}} -2\right)  \right ] \,\eta_{sl} \, I_0 (x,\,0,\,0)~,
\label{gamsl1}
\eeq
and for the non-leptonic decay rate
\beq
\Gamma_{nl}(H_b)={|V_{cb}|^2 \, G_F^2 \, m_b^5 \over 64 \, \pi^3} \,
\langle H_b | {\overline b} b | H_b \rangle \, \left \{\left [1+
 {{\mu_g^2} \over m_b^2} \left ( {x \over 2}
\,{ d \over
{dx}} -2\right)  \right ] \,\eta_{nl} \, I (x)-8 \eta_2 \, {{\mu_g^2}
\over m_b^2} \, I_2(x) \right \}~.
\label{gamnl1}
\eeq
These formulas take into account only the dominant CKM mixing $V_{cb}$
and neglect the small one, $V_{ub}$. The following notation is also
used: $x=m_c/m_b$,  $I_0(x,y,z)$ stands for the kinematical suppression
factor in a three-body weak decay due to masses of the final fermions.
In particular,
\begin{eqnarray}
&&I_0(x,0,0)=(1-x^4)(1-8\, x^2+ x^4)-24 \, x^4 \, \ln x~, \\ \nonumber
&&I_0(x,x,0)=(1-14\, x^2-2 \, x^4 -12 x^6)\sqrt{1-4\, x^2} + 24 \,
(1-x^4) \, \ln {1+ \sqrt{1-4\, x^2} \over 1- \sqrt{1-4\, x^2}}~.
\label{kints}
\end{eqnarray}
Furthermore, $I(x)=I_0(x,0,0)+I_0(x,x,0)$, and
$$
I_2(x)=(1-x^2)^3+\left ( 1+{1 \over
2} x^2+ 3 x^4 \right ) \, \sqrt{1-4\, x^2}-3x^2 \, (1-2x^4)
\ln{{1+\sqrt{1-4\, x^2}} \over {1-\sqrt{1-4\, x^2}}}~.
$$
Finally, the QCD radiative correction factor $\eta_2$ in
eq.(\ref{gamnl1}) is known in the leading logarithmic approximation and
is expressed in terms of the well known coefficients $C_+$ and $C_-$ in
the renormalization of the non-leptonic weak interaction:
$\eta_2=(C_+^2(m_b)-C_-^2(m_b))/6$ with
\beq
C_-(\mu)=C_+^{-2}(\mu)=\left [ {\alpha_s(\mu) \over \alpha_s(m_W)}
\right ]^{4/b}~,
\label{cpm}
\eeq
and $b$ is the coefficient in the QCD beta function. The value of $b$
relevant to $b$ decays is $b=23/3$.

Numerically, for $x \approx 0.3$, the expressions for the decay rates
can be written as
\begin{eqnarray}
&&\Gamma_{sl}(H_b)=\Gamma_{sl}^{parton} \, \left ( 1-
{\mu_\pi^2(H_b)-\mu_g^2(H_b) \over 2 \, m_b^2}-2.6 {\mu_g^2(H_b) \over
m_b^2} \right )~, \nonumber \\
&&\Gamma_{nl}(H_b)=\Gamma_{nl}^{parton} \, \left ( 1-
{\mu_\pi^2(H_b)-\mu_g^2(H_b) \over 2 \, m_b^2}-1.0 {\mu_g^2(H_b) \over
m_b^2} \right )~,
\label{gamnum}
\end{eqnarray}
where $\Gamma^{parton}$ is the perturbation theory value of the
corresponding decay rate of $b$ quark.

The matrix elements $\mu_\pi^2$ and $\mu_g^2$ are related to the
spectroscopic formula for a heavy hadron mass $M$,
\beq
M(H_Q)=m_Q + {\overline \Lambda}(H_Q)+{\mu_\pi^2(H_Q)-\mu_g^2(H_Q) \over
2 \, m_Q} + \ldots
\label{massf}
\eeq
Being combined with the spin counting for pseudoscalar and vector
mesons, this formula allows to find the value of $\mu_g^2$ in
pseudoscalar mesons from the mass splitting:
\beq
\mu_g^2(B)={3 \over 4} \left ( M_{B^*}^2-M_B^2 \right ) \approx 0.36 \,
GeV^2~.
\label{mugn}
\eeq
The value of $\mu_\pi^2$ for $B$ mesons is less certain. It is
constrained by the inequality \cite{mv:95}, $\mu_\pi^2(H_Q) \ge
\mu_g^2(H_Q)$, and there are theoretical estimates from the QCD sum
rules \cite{bb:94}: $\mu_\pi^2(B) = 0.54 \pm 0.12 \, GeV^2$ and from an
analysis of spectroscopy of heavy hadrons \cite{neubert:00}:
$\mu_\pi^2(B) = 0.3 \pm 0.2 \, GeV^2$. In any event, the discussed
corrections are rather small for $b$ hadrons, given that $\mu_g^2/m_b^2
\approx 0.015$. The largest, in relative terms, effect of these
corrections in $B$ meson decays is on the semileptonic decay rate, where
it amounts to 4 -- 5 \% suppression of the rate, which rate however is
only a moderate fraction of the total width. In the dominant
non-leptonic decay rate the effect is smaller, and, according to the
formula (\ref{gamnum}) amounts  to about 1.5 -- 2 \%.

The effect of the $m_Q^{-2}$ corrections can be evaluated with a
somewhat better certainty for the ratio of the decay rates of
$\Lambda_b$ and $B$ mesons. This is due to the fact that
$\mu_g^2(\Lambda_b)=0$, since there is no correlation of the spin of the
heavy quark in $\Lambda_b$ with the light component, having overall
quantum numbers $J^P=0^+$. Then, applying the formula (\ref{gamnum}) to
$B$ and $\Lambda_b$, we find for the ratio of the (dominant)
non-leptonic decay rates:
\beq
{\Gamma_{nl}(\Lambda_b) \over \Gamma_{nl}(B)} = 1
-{\mu_\pi^2(\Lambda_b)-\mu_\pi^2(B) \over 2 \, m_b^2}+0.5 {\mu_g^2(B)
\over m_b^2}~.
\label{lambdab}
\eeq
The difference of the kinetic terms,
$\mu_\pi^2(\Lambda_b)-\mu_\pi^2(B)$, can be estimated from the mass
formula:
\beq
\mu_\pi^2(\Lambda_b)-\mu_\pi^2(B)={2 \, m_b \, m_c \over m_b-m_c} \left
[ {\overline M}(B)-{\overline M}(D)-M(\Lambda_b)+ M(\Lambda_c) \right] =
0 \pm 0.04 \, GeV^2~,
\label{difmu}
\eeq
where ${\overline M}$ is the spin-averaged mass of the mesons, e.g.
${\overline M}(B)=(M(B)+3 M(B^*))/4$. The estimated difference of the
kinetic terms is remarkably small. Thus the effect in the ratio of
the decay rates essentially reduces to the chromomagnetic term, which is
also rather small and accounts for less than 1\% difference of the
rates. For the ratio of the semileptonic decay rates the chromomagnetic
term is approximately four times larger, but then the contribution of
the semileptonic rates to the total width is rather small. Thus one
concludes that the terms of order $m_b^{-2}$ in the OPE expansion for
the decay rates can account only for about 1\% difference
of the lifetimes of $\Lambda_b$ and the $B$ mesons.

The significance of the $m_Q^{-2}$ terms is substantially different for
the decay rates of charmed hadrons, where these effects suppress the
inclusive decays of the $D$ mesons by about 40\% with respect to those
of the charmed hyperons in a reasonable agreement with the observed
pattern of the lifetimes.

It should be emphasized once again that the $m_Q^{-2}$ effects do not
depend on the flavors of the spectator quarks or antiquarks. Thus the
explanation of the variety of the inclusive decay rates within the
flavor SU(3) multiplets, observed for charmed hadrons and expected for
the $b$ ones, has to be sought among the $m_Q^{-3}$ terms.

\section{$L_{eff}^{(3)}$. Coefficients and operators}

Although the third term in the expansion (\ref{first3}) is formally
suppressed by an extra power of $m_Q^{-1}$, its effects are comparable
to, or even larger than the effects of the second term. This is due to
the fact that the diagrams determining the third term (see Fig.
1) contain a two-body phase space, while the first two terms
involve a three-body phase space. This brings in a numerical enhancement
factor, typically $4 \pi^2$. The enhanced numerical significance of the
third term in OPE, generally, does not signal a poor convergence of the
expansion in inverse heavy quark mass for decays of $b$, and even
charmed, hadrons the numerical enhancement factor is a one time
occurrence in the series, and there is no reason for similar `anomalous'
enhancement among the higher terms in the expansion.

Here we first present the expressions for the relevant parts of
$L_{eff}^{(3)}$ for decays of $b$ and $c$ hadrons in the form of
four-quark operators and then proceed to a discussion of hadronic matrix
elements and the effects in specific inclusive decay rates. The
consideration of the effects in decays of charmed hadrons is interesting
in its own right, and leads to new predictions to be tested
experimentally, and is also important for understanding the magnitude of
the involved matrix elements using the existing data on charm decays.

We start with considering the term $L_{eff}^{(3)}$ in $b$ hadron
non-leptonic decays, $L_{eff, nl}^{(3,b)}$, induced by the underlying
processes $b \to c \, {\overline u} \, d$, $b \to c \, {\overline c} \,
s$, $b \to c \, {\overline u} \, s$, and $b \to c \, {\overline c} \,
d$. Unlike the case of three-body decay, the kinematical difference
between the two-body states $c \overline c$ and $c \overline u$,
involved in calculation of $L_{eff, nl}^{(3,b)}$ is of the order of
$m_c^2/m_b^2 \approx 0.1$ and is rather small. At present level of
accuracy in discussing this term in OPE, one can safely neglect the
effect of finite charmed quark mass\footnote{The full expression for a
finite charmed quark mass can be found in \cite{ns:97}}. In this
approximation the expression for $L_{eff, nl}^{(3,b)}$ reads as
\cite{sv:86}
\begin{eqnarray}
\label{l3nlb}
&&L_{eff, \, nl}^{(3, b)}=  |V_{bc}|^2 \,{G_F^2 \, m_b^2 \over 4
\pi} \,
\left \{
{\tilde C}_1 \, (\overline b \Gamma_\mu b)(\overline u \Gamma_\mu u) +
{\tilde C}_2  \,
(\overline b \Gamma_\mu u) (\overline u \Gamma_\mu b) +\right .
\nonumber \\
&& {\tilde C}_5 \, (\overline  b \Gamma_\mu b +
{2 \over 3}\overline b \gamma_\mu \gamma_5 b) (\overline q \Gamma_\mu
q)+ {\tilde C}_6 \, (\overline  b_i \Gamma_\mu b_k +
{2 \over 3}\overline b_i \gamma_\mu \gamma_5 b_k)
(\overline q_k \Gamma_\mu q_i)+ \\ \nonumber
&&\left. {1 \over 3} \,
{\tilde \kappa}^{1/2} \, ({\tilde \kappa}^{-2/9}-1) \, \left [ 2 \,
({\tilde C}_+^2 - {\tilde C}_-^2) \,
 (\overline b \Gamma_\mu t^a b) \,
j_\mu^a - \right. \right. \\ \nonumber
&& \left . \left.
(5{\tilde C}_+^2+{\tilde C}_-^2 - 6 \, {\tilde C}_+ \, {\tilde C}_-)
(\overline  b \Gamma_\mu t^a b +
{2 \over 3}\overline b \gamma_\mu \gamma_5 t^a b) j_\mu^a \right ]
\right \}~,
\end{eqnarray}
where the notation $(\overline q \, \Gamma \, q)= (\overline d \,
\Gamma \, d) + (\overline s \, \Gamma \, s)$ is used, the indices $i, \,
k$ are the color triplet ones, $\Gamma_\mu=\gamma_\mu \, (1-\gamma_5)$,
and $j_\mu^a=\overline u \gamma_\mu t^a u + \overline d \gamma_\mu t^a d
+ \overline s \gamma_\mu t^a s$ is the color current of the light quarks
with $t^a = \lambda^a /2$ being the generators of the color SU(3). The
notation ${\tilde C}_\pm$, is used as shorthand for the short-distance
renormalization coefficients $C_\pm(\mu)$ at $\mu=m_b$: ${\tilde C}_\pm
\equiv C_\pm(m_b)$. The expression (\ref{l3nlb}) is written in the
leading logarithmic approximation for the QCD radiative effects in a low
normalization point $\mu$ such that $\mu \ll m_b$ (but still, at least
formally, $\mu \gg \Lambda_{QCD}$). For such $\mu$ there arises so
called `hybrid' renormalization \cite{sv:87}, depending on the factor
${\tilde \kappa}=\alpha_s(\mu)/\alpha_s(m_b)$. The coefficients ${\tilde
C}_A$ with $A=1, \ldots , 6$ in eq.(\ref{l3nlb}) have the following
explicit expressions in terms of ${\tilde C}_\pm$ and ${\tilde \kappa}$:
\begin{eqnarray}
&&{\tilde C}_1= {\tilde C}_+^2+{\tilde C}_-^2 + {1 \over 3} (1 -
\kappa^{1/2}) ({\tilde C}_+^2-{\tilde C}_-^2)~,
\nonumber \\
&&{\tilde C}_2= \kappa^{1/2} \, ({\tilde C}_+^2-{\tilde C}_-^2)~,
\nonumber \\
&&{\tilde C}_3=- {1 \over 4} \, \left [ ({\tilde C}_+-{\tilde C}_-)^2 +
{1 \over 3}
(1-\kappa^{1/2})
(5{\tilde C}_+^2+{\tilde C}_-^2+6{\tilde C}_+{\tilde C}_-) \right] ~,
\nonumber \\
&&{\tilde C}_4=-{1 \over 4} \, \kappa^{1/2} \, (5{\tilde C}_+^2+{\tilde
C}_-^2+6{\tilde C}_+{\tilde C}_-)~, \nonumber
\\
&&{\tilde C}_5=-{1 \over 4} \, \left [ ({\tilde C}_++{\tilde C}_-)^2 +
{1 \over 3} (1-\kappa^{1/2})
(5{\tilde C}_+^2+{\tilde C}_-^2-6{\tilde C}_+{\tilde C}_-) \right]~,
\nonumber \\
&&{\tilde C}_6=-{1 \over 4} \, \kappa^{1/2} \, (5{\tilde C}_+^2+{\tilde
C}_-^2-6{\tilde C}_+{\tilde C}_-)~.
\label{coefs}
\end{eqnarray}

The expression for the CKM dominant semileptonic decays of $b$ hadrons,
associated with the elementary process $b \to c \, \ell \, \nu$ does not
look to be of an immediate interest. The reason is that this process is
intrinsically symmetric under the flavor SU(3), and one expects no
significant splitting of the semileptonic decay rates within SU(3)
multiplets of the $b$ hadrons. The only possible effect of this term,
arising through a penguin-like mechanism can be in a small overall
shift
of semileptonic decay rates between $B$ mesons and baryons. However,
these effects are quite suppressed and are believed to be even smaller
than the ones arising form the discussed $m_b^{-2}$ terms.

For charm decays there is a larger, than for $b$ hadrons, variety of
effects associated with $L_{eff}^{(3)}$, that can be studied
experimentally, and we present here the relevant parts of the effective
Lagrangian. For the CKM dominant non-leptonic decays of charm,
originating from the quark process $c \to s \, u \, {\overline d}$, the
discussed term in OPE has the form
\begin{eqnarray}
\label{l3nl}
&&L_{eff, nl }^{(3, \, \Delta C = \Delta S)}= \cos^4\theta_c \,{G_F^2 \,
m_c^2 \over 4 \pi} \,
\left \{
C_1 \, (\overline c \Gamma_\mu c)(\overline d \Gamma_\mu d) + C_2  \,
(\overline c \Gamma_\mu d) (\overline d \Gamma_\mu c) +\right .
\nonumber \\
&& C_3 \, (\overline  c \Gamma_\mu c +
{2 \over 3}\overline c \gamma_\mu \gamma_5 c) (\overline s \Gamma_\mu
s)+ C_4 \, (\overline  c_i \Gamma_\mu c_k +
{2 \over 3}\overline c_i \gamma_\mu \gamma_5 c_k)
(\overline s_k \Gamma_\mu s_i) +
\\ \nonumber
&& C_5 \, (\overline  c \Gamma_\mu c +
{2 \over 3}\overline c \gamma_\mu \gamma_5 c) (\overline u \Gamma_\mu
u)+ C_6 \, (\overline  c_i \Gamma_\mu c_k +
{2 \over 3}\overline c_i \gamma_\mu \gamma_5 c_k)
(\overline u_k \Gamma_\mu u_i)+ \\ \nonumber
&&\left. {1 \over 3} \,
\kappa^{1/2} \, (\kappa^{-2/9}-1) \, \left [ 2 \, (C_+^2 - C_-^2) \,
 (\overline c \Gamma_\mu t^a c) \,
j_\mu^a - (5C_+^2+C_-^2)
(\overline  c \Gamma_\mu t^a c +
{2 \over 3}\overline c \gamma_\mu \gamma_5 t^a c) j_\mu^a \right ]
\right \}~,
\end{eqnarray}
where, $\theta_c$ is the Cabibbo angle, and the coefficients without the
tilde are given by the same expressions as above for the $b$ decays
(i.e. those with tilde) with the replacement $m_b \to m_c$. The part of
the notation in the superscript $\Delta C = \Delta S$ points to the
selection rule for the dominant CKM unsuppressed non-leptonic decays.
One can rather realistically envisage however a future study of
inclusive rates for the once CKM suppressed decays of charmed
hadrons\footnote{Even if the inclusive rate of these decays is not to be
separated experimentally, they contribute about 10\% of the total decay
rate, and it is worthwhile to include their contribution in the balance
of the total width.}, satisfying the selection rule $\Delta S=0$ and
associated with the quark processes $c \to d \, u \, {\overline s}$ and
$c \to d \, u \,
{\overline d}\,$. The corresponding part of the effective Lagrangian for
these processes reads as
\begin{eqnarray}
&&L_{eff, \, nl}^{(3, \Delta S=0)}= \cos^2 \theta_c \, \sin^2 \theta_c
\,{G_F^2 \, m_c^2 \over 4 \pi} \,
\left \{
C_1 \, (\overline c \Gamma_\mu c)(\overline q \Gamma_\mu q) + C_2  \,
(\overline c_i \Gamma_\mu c_k) (\overline q_k \Gamma_\mu q_i) +\right .
\nonumber \\
&& C_3 \, (\overline  c \Gamma_\mu c +
{2 \over 3}\overline c \gamma_\mu \gamma_5 c) (\overline q \Gamma_\mu
q)+ C_4 \, (\overline  c_i \Gamma_\mu c_k +
{2 \over 3}\overline c_i \gamma_\mu \gamma_5 c_k)
(\overline q_k \Gamma_\mu q_i) +  \\ \nonumber
&& 2 \, C_5 \, (\overline  c \Gamma_\mu c +
{2 \over 3}\overline c \gamma_\mu \gamma_5 c) (\overline u \Gamma_\mu
u)+ 2 \, C_6 \, (\overline  c_i \Gamma_\mu c_k +
{2 \over 3}\overline c_i \gamma_\mu \gamma_5 c_k)
(\overline u_k \Gamma_\mu u_i)+ \\ \nonumber
&&\left. {2 \over 3} \,
\kappa^{1/2} \, (\kappa^{-2/9}-1) \, \left [ 2 \, (C_+^2 - C_-^2) \,
 (\overline c \Gamma_\mu t^a c) \,
j_\mu^a - (5C_+^2+C_-^2)
(\overline  c \Gamma_\mu t^a c +
{2 \over 3}\overline c \gamma_\mu \gamma_5 t^a c) j_\mu^a \right ]
\right \}~
\label{l3nl1}
\end{eqnarray}
where again the notation $(\overline q \, \Gamma \, q)= (\overline d \,
\Gamma \, d) + (\overline s \, \Gamma \, s)$ is used.

The semileptonic decays of charm, the CKM dominant, associated with $c
\to s \, \ell \, \nu$, and the CKM suppressed, originating from $c \to s
\, \ell \, \nu$, contribute to the semileptonic decay rate, which
certainly can be measured experimentally. The expression for the part of
the effective Lagrangian, describing the $m_Q^{-3}$ terms in these
decays is \cite{mv:96,cheng:97,gm:98}
\begin{eqnarray}
&&L_{eff, \, sl}^{(3)}= \nonumber \\
&&{G_F^2 \, m_c^2 \over 12 \pi} \, \left \{  \cos^2 \theta_c \, \left [
L_1 \, (\overline  c \Gamma_\mu c +
{2 \over 3}\overline c \gamma_\mu \gamma_5 c) (\overline s \Gamma_\mu
s)+ L_2 \, (\overline  c_i \Gamma_\mu c_k +  {2 \over 3}\overline c_i
\gamma_\mu \gamma_5 c_k)
(\overline s_k \Gamma_\mu s_i) \right ] + \right. \nonumber \\
&& \sin^2 \theta_c \, \left [
L_1\, (\overline  c \Gamma_\mu c +
{2 \over 3}\overline c \gamma_\mu \gamma_5 c) (\overline d \Gamma_\mu
d)+L_2 \, (\overline  c_i \Gamma_\mu c_k +
{2 \over 3}\overline c_i \gamma_\mu \gamma_5 c_k)
(\overline d_k \Gamma_\mu d_i) \right ] -  \nonumber \\
&& \left. 2 \, \kappa^{1/2} \, (\kappa^{-2/9}-1) \,
(\overline  c \Gamma_\mu t^a c +
{2 \over 3}\overline c \gamma_\mu \gamma_5 t^a c) j_\mu^a
\right \} ~,
\label{l3sl}
\end{eqnarray}
with the coefficients $L_1$ and $L_2$ found as
\beq
L_1=(\kappa^{1/2}-1), ~~~~~
L_2 = -   3\, \kappa^{1/2}~.
\label{coefl}
\eeq

\section{Effects of $L_{eff}^{(3)}$ in mesons}

The expressions for the terms in $L_{eff}^{(3)}$ still leave us with the
problem of evaluating the matrix elements of the four-quark operators
over
heavy hadrons in order to calculate the effects in the decay rates
according to the formula (\ref{lgam}). In doing so only few
conclusions can be drawn in a reasonably model independent way, i.e.
without resorting to evaluation of the matrix elements using specific
ideas about the dynamics of quarks inside hadrons. The most
straightforward prediction can in fact be found for $b$ hadrons. Namely,
one can notice that the operator (\ref{l3nlb}) is symmetric under the
flavor U spin (an SU(2) subgroup of the flavor SU(3), which mixes $s$
and $d$ quarks). This is a direct consequence of neglecting the small
kinematical effect of the charmed quark mass. However the usual
(in)accuracy of the flavor SU(3) symmetry is likely to be a more
limiting factor for the accuracy of this symmetry, than the corrections
of order $m_c^2/m_b^2$. Modulo this reservation the immediate prediction
from this symmetry is the degeneracy of inclusive decay rates within U
spin doublets:
\beq
\Gamma(B_d)=\Gamma(B_s)\,,~~~~\Gamma(\Lambda_b)=\Gamma(\Xi_b^0)~,
\label{upred}
\eeq
where $\Gamma(B_s)$ stands for the average rate over the two eigenstates
of the $B_s - {\overline B}_s$ oscillations. The data on decay rates of
the cascade hyperon $\Xi_b^0$ are not yet available, while the currently
measured lifetimes of $B_d$ and $B_s$ are within less than 2\% from one
another. Theoretically, the difference of the lifetimes, associated with
possible violation of the SU(3) symmetry and with breaking of the U
symmetry of the effective Lagrangian (\ref{l3nlb}), is expected to not
exceed about 1\%.

For the non-vanishing matrix elements of four-quark operators over
pseudoscalar mesons one traditionally starts with the factorization
formula and parametrizes possible deviation from factorization in terms
of `bag constants'. Within the normalization convention adopted here the
relations used in this parametrization read as
\begin{eqnarray}
&&\langle P_{Q \overline q} | (\overline Q \Gamma_\mu q) \, (\overline q
\Gamma_\mu Q) | P_{Q \overline q} \rangle = {1 \over 2} \, f_P^2 \, M_P
\,B \,, \nonumber \\
&&\langle P_{Q \overline q} | (\overline Q \Gamma_\mu Q) \, (\overline q
\Gamma_\mu q) | P_{Q \overline q} \rangle = {1 \over 6} \, f_P^2 \, M_P
\,{\tilde B}\,,
\label{bagc}
\end{eqnarray}
where $P_{Q \overline q}$ stands for pseudoscalar meson made of $Q$ and
${\overline q}$, $f_P$ is the annihilation constant for the meson, and
$B$ and ${\tilde B}$ are bag constants. The parameters $B$ and ${\tilde
B}$ generally depend on the normalization point $\mu$ for the operators,
and this dependence is compensated by the $\mu$ dependence of the
coefficients in $L_{eff}^{(3)}$, so that the results for the physical
decay rate difference do not depend on $\mu$. If the normalization point
$\mu$ is chosen at the heavy quark mass (i.e. $\mu=m_b$ for $B$ mesons,
and $\mu=m_c$ for $D$ mesons) the predictions for the difference of
total
decay rates take a simple form in terms of the corresponding bag
constants (generally different between $B$ and $D$):
\begin{eqnarray}
\Gamma(B^\pm) - \Gamma(B^0) &=& |V_{cb}|^2 \,{G_F^2 \, m_b^3 \, f_B^2
\over 8
\pi} \left [ ({\tilde C}_+^2-{\tilde C}_-^2) \, B(m_b) + {1 \over 3}\,
({\tilde C}_+^2+{\tilde C}_-^2) \,  {\tilde B}(m_b) \right] \nonumber \\
&\approx& - 0.025 \, \left ( {f_B \over 200 \, MeV } \right )^2 \,
ps^{-1}~,
\label{bpred}
\end{eqnarray}
\begin{eqnarray}
\Gamma(D^\pm) - \Gamma(D^0) &=&  \cos^4 \theta_c \,{G_F^2 \, m_c^3 \,
f_D^2 \over 8
\pi} \left [ (C_+^2-C_-^2) \,  B(m_c) + {1 \over 3}\, (C_+^2+C_-^2) \,
{\tilde B}(m_c) \right] \nonumber \\
&\sim& - 0.8 \, \left ( {f_D \over 200 \, MeV } \right )^2 \, ps^{-1}~,
\label{dpred}
\end{eqnarray}
where the numerical values are written in the approximation of exact
factorization: $B=1$, and ${\tilde B}=1$. It is seen from the numerical
estimates that, even given all the theoretical uncertainties, the
presented approach is in reasonable agreement with the data on the
lifetimes of $D$ and $B$ mesons. In particular, this approach describes,
at least qualitatively, the strong suppression of the decay rate of
$D^\pm$ mesons relative to $D^0$, the experimental observation of which
has in fact triggered in early 80-s the theoretical study of
preasymptotic in heavy quark mass effects in inclusive decay rates.
For the $B$ mesons the estimate (\ref{bpred}) is also in a reasonable
agreement with the current data for the discussed difference ($-0.043
\pm 0.017 \, ps^{-1}$).

\section{Effects of $L_{eff}^{(3)}$ in baryons}

The weakly decaying heavy hyperons, containing either $c$ or $b$ quark
are: $\Lambda_Q \sim Qud,~ \Xi_Q^{(u)} \sim Qus,~ \Xi_Q^{(d)} \sim Qds$,
and $\Omega_Q \sim Qss$. The first three baryons form an SU(3)
(anti)triplet. The light diquark in all three is in the state with
quantum numbers $J^P = 0^+$, so that there is no correlation of the spin
of the heavy quark with the light component of the baryon. On the
contrary, in $\Omega_Q$ the two strange quarks form a $J^P = 1^+$ state,
and a correlation between the spins of heavy and light quarks is
present. The absence of spin correlation for the heavy quark in the
triplet of hyperons somewhat reduces the number of independent
four-quark operators, having nonvanishing diagonal matrix elements over
these baryons. Indeed, the operators entering $L_{eff}^{(3)}$ contain
both vector and axial bilinear forms for the heavy quarks. However the
axial part requires a correlation of the heavy quark spin with that of a
light quark, and is thus vanishing for the hyperons in the triplet.
Therefore only the structures with vector currents are relevant for
these hyperons. These structures are of the type $(\overline c \,
\gamma_\mu \, c)
(\overline q \, \gamma_\mu \, q)$ and $(\overline c_i \, \gamma_\mu \,
c_k)
(\overline q_k \, \gamma_\mu \, q_i)$ with $q$ being $d$, $s$ or $u$.
The flavor SU(3) symmetry then allows to express, for each of the two
color combinations, the matrix elements of three different operators,
corresponding to three flavors of $q$, over the baryons in the triplet
in terms of only two combinations: flavor octet and flavor singlet. Thus
all effects of $L_{eff}^{(3)}$ in the triplet of the baryons can be
expressed in terms of four independent combinations of matrix elements.
These can be chosen in the following way:
\begin{eqnarray}
\label{defxy}
x&=&\left \langle  {1 \over 2} \, (\overline Q \, \gamma_\mu \, Q) \left
[
(\overline u \, \gamma_\mu u) - (\overline s \, \gamma_\mu s) \right]
\right \rangle_{\Xi_Q^{(d)}-\Lambda_Q} = \left \langle  {1 \over 2} \,
(\overline Q \, \gamma_\mu \, Q) \left [ (\overline s \, \gamma_\mu s) -
(\overline d \, \gamma_\mu d) \right] \right \rangle_{\Lambda_Q -
\Xi_Q^{(u)}}~,  \\ \nonumber
y&=&\left \langle  {1 \over 2} \, (\overline Q_i \, \gamma_\mu \, Q_k)
\left [ (\overline u_k \, \gamma_\mu u_i) - (\overline s_k \, \gamma_\mu
s_i) \right ] \right \rangle_{\Xi_Q^{(d)}-\Lambda_Q} = \left \langle  {1
\over 2} \, (\overline Q_i \, \gamma_\mu \, Q_k) \left [ (\overline s_k
\, \gamma_\mu s_i) - (\overline d_k \, \gamma_\mu d_i) \right] \right
\rangle_{\Lambda_Q - \Xi_Q^{(u)}}~,
\end{eqnarray}
with the notation for the differences of the matrix elements:
$\langle {\cal O} \rangle_{A-B}= \langle A | {\cal O} | A \rangle -
\langle B | {\cal O} | B \rangle$, for the flavor octet part and
the matrix elements:
\begin{eqnarray}
&&x_s = {1 \over 3} \, \langle H_Q | ({\overline Q} \, \gamma_\mu \, Q)
\left ( ({\overline u} \, \gamma_\mu \, u)+({\overline d} \, \gamma_\mu
\, d)+ ({\overline s} \, \gamma_\mu \, s) \right ) | H_Q \rangle
\nonumber \\
&&y_s = {1 \over 3} \, \langle H_Q | ({\overline Q}_i \, \gamma_\mu \,
Q_k) \left ( ({\overline u}_k \, \gamma_\mu \, u_i)+({\overline d}_k \,
\gamma_\mu \, d_i)+ ({\overline s}_k \, \gamma_\mu \, s_i) \right ) |
H_Q \rangle
\label{xys}
\end{eqnarray}
for the flavor singlet part, where $H_Q$ stands for any heavy hyperon in
the (anti)triplet.

The initial, very approximate, theoretical estimates of the matrix
elements \cite{sv:86} were essentially based on a non-relativistic
constituent quark model, where these matrix elements are proportional to
the density of a light quark at the location of the heavy one, i.e. in
terms of the wave function, proportional to $|\psi(0)|^2$. Using then
the same picture for the matrix elements over pseudoscalar mesons,
relating the quantity $|\psi(0)|^2$ to the annihilation constant $f_P$,
and assuming that $|\psi(0)|^2$ is approximately the same in baryons as
in mesons, one arrived at the estimate
\beq
y=-x=x_s=-y_s \approx {f_D^2 \, M_D \over 12} \approx 0.006 \, GeV^2~,
\label{xyold}
\eeq
where the sign relation between $x$ and $y$ is inferred from the color
antisymmetry of the constituent quark wave function for baryons. Since
the constituent picture was believed to be valid at distances of the
order of the hadron size, the estimate (\ref{xyold}) was applied to the
matrix elements in a low normalization point where $\alpha_s(\mu)
\approx 1$. For the matrix elements of the operators, containing $s$
quarks over the $\Omega_Q$ hyperon, this picture predicts an
enhancement factor due to the spin correlation:
\beq
\langle \Omega_Q | ({\overline Q} \, \Gamma_\mu \, Q)
 ({\overline s} \, \Gamma_\mu \, s)  |\Omega_Q \rangle = - \langle
\Omega_Q | ({\overline Q}_i \, \Gamma_\mu \, Q_k)
 ({\overline s}_k \, \Gamma_\mu \, s_i)  |\Omega_Q \rangle = {10 \over
3}\, y
\label{omegaold}
\eeq
Although these simple estimates allowed to correctly predict
\cite{sv:86} the hierarchy of lifetimes of charmed hadrons prior to
establishing this hierarchy experimentally, they fail to quantitatively
predict the differences of lifetimes of charmed baryons. We shall see
that the available data indicate that the color antisymmetry relation is
badly broken, and the absolute value of the matrix elements is larger,
than the naive estimate (\ref{xyold}), especially for the quantity $x$.

It should be emphasized that in the heavy quark limit the matrix
elements (\ref{defxy}) and (\ref{xys}) do not depend on the flavor of
the heavy quark, provided that the same normalization point $\mu$ is
used. Therefore, applying the OPE formulas to both charmed and $b$
baryons, one can extract the values for the matrix elements from
available data on charmed hadrons, and then make predictions for $b$
baryons, as well as for other inclusive decay rates, e.g. semileptonic,
for charmed hyperons.

The only data available so far, which would allow to extract the matrix
elements, are on the lifetimes of charmed hyperons. Therefore, one has
to take into account several essential types of inclusive decay,  at
least those that contribute to the total decay rate at the level of
about 10\%. Here we first concentrate on the differences of the decay
rates within the SU(3) triplet of the hyperons, which will allow us to
extract the non-singlet quantities $x$ and $y$, and then discuss the
SU(3) singlet shifts of the rates.

The differences of the dominant Cabibbo unsuppressed
non-leptonic decay rates are given by
\begin{eqnarray}
\delta_1^{nl, \,0} \equiv \Gamma^{nl}_{\Delta S =
\Delta C}(\Xi_c^0)-\Gamma^{nl}_{\Delta S = \Delta C}(\Lambda_c) = \cos^4
\theta_c \, {G_F^2 \,
m_c^2 \over 4 \pi} \left [ (C_5 - C_3) \, x + (C_6 - C_4) \, y \right
]~, \nonumber \\
\delta_2^{nl, \,0} \equiv \Gamma^{nl}_{\Delta S =
\Delta C}(\Lambda_c)-\Gamma^{nl}_{\Delta S =\Delta C}(\Xi_c^+) = \cos^4
\theta_c \, {G_F^2 \,
m_c^2 \over 4 \pi} \left [ (C_3 - C_1) \, x + (C_4 - C_2) \, y \right
]~.
\label{dnl0}
\end{eqnarray}
The once Cabibbo suppressed decay rates of $\Lambda_c$ and $\Xi_c^+$ are
equal, due to the $\Delta U =0$ property of the corresponding effective
Lagrangian $L_{eff, nl}^{(3,1)}$ (eq.(\ref{l3nl1})). Thus the only
difference for these decays in the baryon triplet is
\beq
\delta^{nl,1} \equiv \Gamma^{nl}_{\Delta S =0}
(\Xi_c^0)-\Gamma^{nl}_{\Delta S = 0 }(\Lambda_c) = \cos^2 \theta_c \,
\sin^2 \theta_c \, {G_F^2
\, m_c^2 \over 4 \pi} \left [ (2\, C_5 - C_1 - C_3) \, x + (2 \, C_6 -
C_2 - C_4) \, y \right ]~.
\label{dnl1}
\eeq
The dominant semileptonic decay rates are equal among the two $\Xi_c$
baryons due to the isotopic spin property $\Delta I =0$ of the
corresponding interaction Lagrangian, thus there is only one non-trivial
splitting for these decays:
\beq
\delta^{sl,0} \equiv \Gamma^{sl}_{\Delta S = -1}(\Xi_c) -
\Gamma^{sl}_{\Delta S = -1} (\Lambda_c) = - \cos^2 \theta_c {G_F^2 \,
m_c^2 \over 12
\pi} \left [ L_1 \, x + L_2 \, y \right ]~.
\label{dsl0}
\eeq
Finally, the Cabibbo suppressed semileptonic decay rates are equal for
$\Lambda_c$ and $\Xi_c^0$, due to the $\Delta V =0$ property of the
corresponding interaction. Thus the only difference for these is
\beq
\delta^{sl,1} \equiv \Gamma^{sl}_{\Delta S = 0}(\Lambda_c) -
\Gamma^{sl}_{\Delta S = 0}(\Xi_c^+) = - \sin^2 \theta_c {G_F^2 \, m_c^2
\over 12
\pi} \left [ L_1 \, x + L_2 \, y \right ]~.
\label{dsl1}
\eeq

Using the relations (\ref{dnl0}) - (\ref{dsl1}) on can find expressions
for two differences of the measured total decay rates, $\Delta_1 =
\Gamma (\Xi_c^0) - \Gamma (\Lambda_c)$ and $\Delta_2= \Gamma
(\Lambda_c) - \Gamma (\Xi_c^+)$, in terms of the quantities $x$ and $y$:
\begin{eqnarray}
&\Delta_1= \delta_1^{nl, \,0} + \delta^{nl,1} + 2 \, \delta^{sl,0}=
\nonumber \\
&{G_F^2 \, m_c^2 \over 4 \pi} \, \cos^2 \theta \, \left \{ x
\, \left [ \cos^2 \theta \, (C_5-C_3) + \sin^2 \theta \, (2 \, C_5 - C_1
-C_3)- {2 \over 3} \, L_1 \right ] + \right.
\nonumber \\
&\left. y \, \left[ \cos^2 \theta \,
(C_6-C_4) + \sin^2 \theta \, (2 \, C_6 - C_2 -C_4)- {2 \over 3} \,
L_2\right ] \right \}~,
\label{d1}
\end{eqnarray}
and
\begin{eqnarray}
&\Delta_2= \delta_2^{nl, \,0} - 2 \, \delta^{sl,0} + 2 \, \delta^{sl,1}=
\nonumber \\
&{G_F^2 \, m_c^2 \over 4 \pi} \ \left \{ x \left[ \cos^4
\theta \, (C_3-C_1) + {2 \over 3} \, (\cos^2 \theta - \sin^2 \theta) \,
L_1 \right] +
\right.
\nonumber \\
&\left. y \left[ cos^4 \theta \, (C_4-C_2) + {2 \over 3} \,
(\cos^2 \theta - \sin^2 \theta) \, L_2 \right] \right \}~ .
\label{d2}
\end{eqnarray}

By comparing these relations with the data, one can extract the values
of $x$ and $y$. Using the current data for the total decay rates:
$\Gamma(\Lambda_c)=4.85 \pm 0.28 \, ps^{-1}$,
$\Gamma(\Xi_c^0)= 10.2 \pm 2 \, ps^{-1}$, and the updated value
\cite{pdg1} $\Gamma(\Xi_c^+)=3.0 \pm
0.45 \, ps^{-1}$, we find for the $\mu$ independent matrix element $x$
\beq
x = -(0.04 \pm 0.01) \, GeV^3 \, \left ( 1.4 \, GeV \over m_c \right
)^2~,
\label{resx}
\eeq
while the dependence of the thus extracted matrix element $y$ on the
normalization point $\mu$ is shown in Fig. 2 \,\footnote{It
should be
noted that the curves at large values of $\kappa$, $\kappa > \, \sim 3$,
are shown only for illustrative purpose. The coefficients in the OPE,
leading to the equations (\ref{d1},\ref{d2}), are purely perturbative.
Thus,
formally, they correspond to $\alpha_s(\mu) \ll 1$, i.e. to $\kappa \ll
1/\alpha_s(m_c) \sim (3 - 4)$.} .

\begin{figure}[ht]
  \begin{center}
    \leavevmode
    \epsfbox{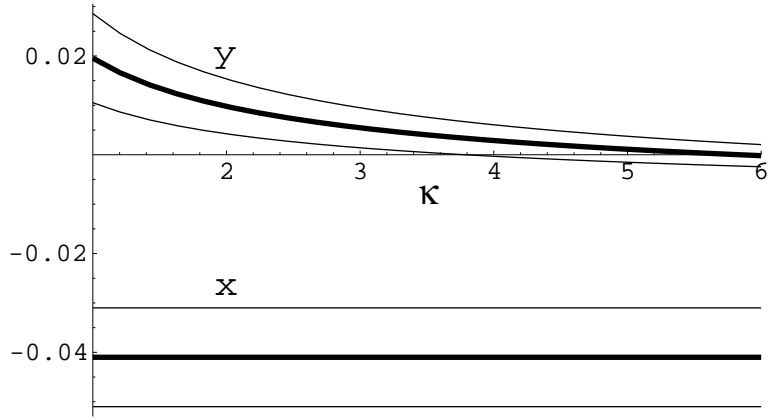}
    \caption{The values of the extracted matrix elements $x$ and $y$ in
$GeV^3$ vs. the normalization point parameter
$\kappa=\alpha_s(\mu)/\alpha_s(m_c)$. The thick lines correspond to the
central value of the data on lifetimes of charmed baryons, and the thin
lines show the error corridors. The extracted values of $x$ and $y$
scale as $m_c^{-2}$ with the assumed mass of the charmed quark, and the
plots are shown for $m_c=1.4 \, GeV$.}
  \end{center}
\label{fig:xy}
\end{figure}

Notably, the extracted values of $x$ and $y$ are in a drastic variance
with the simplistic constituent model: the color antisymmetry relation,
$x=-y$, does not hold at any reasonable $\mu$, and the absolute value of
$x$ is substantially enhanced\footnote{A similar, although with a
smaller enhancement, behavior of the matrix elements was observed in a
recent preliminary lattice study \cite{psm:99}.}

Once the non-singlet matrix elements are determined, they can be used
for predicting differences of other inclusive decay rates within the
triplet of charmed hyperons as well as for the $b$ baryons. Due to
correlation of errors in $x$ and $y$ it makes more sense to express the
predictions directly in terms of the total decay rates of the charmed
hyperons. The thus arising relations between the rates do not depend on
the normalization parameter $\mu$. In this way one finds \cite{mv:99-1}
for the difference of the Cabibbo dominant semileptonic decay rates
between either of the $\Xi_c$ hyperons and $\Lambda_c$:
\beq
\Gamma_{sl}(\Xi_c)-\Gamma_{sl}(\Lambda_c) \approx \delta^{sl, 0}= 0.13
\, \Delta_1 - 0.065 \, \Delta_2 \approx 0.59 \pm 0.32 \, ps^{-1}~.
\label{dslr}
\eeq
When compared with the data on the total semileptonic decay rate of
$\Lambda_c$, $\Gamma_{sl}(\Lambda_c)=0.22 \pm 0.08 \, ps^{-1}$, this
prediction implies that the semileptonic decay rate of the charmed
cascade hyperons can be 2--3 times larger than that of $\Lambda_c$.

The predictions found in a similar way for the inclusive Cabibbo
suppressed decay rates are \cite{mv:99-1}: for non-leptonic decays
\beq
\delta^{nl,1}=0.082 \, \Delta_1 + 0.054 \, \Delta_2 \approx 0.55 \pm
0.22 \, ps^{-1}
\label{dnl1r}
\eeq
and for the semileptonic ones
\beq
\delta^{sl,1}=\tan^2 \theta_c \, \delta^{sl, 0} \approx 0.030 \pm 0.016
\, ps^{-1}~.
\label{dsl1r}
\eeq

For the only difference of the inclusive rates in the triplet of $b$
baryons, $\Gamma(\Lambda_b)-\Gamma(\Xi_b^-)$, one finds an expression in
terms of $x$ and $y$, or alternatively, in terms of the differences
$\Delta_1$ and $\Delta_2$ between the charmed hyperons,
\begin{eqnarray}
&&\Gamma(\Lambda_b)-\Gamma(\Xi_b^-)= \cos^2 \theta_c \, |V_{bc}|^2
\,{G_F^2 \, m_b^2 \over 4 \pi} \, \left[ (\tilde C_5 - \tilde C_1) \, x
+ (\tilde C_6 - \tilde C_2) \, y \right] \approx \nonumber \\
&&|V_{bc}|^2  \, {m_b^2 \over m_c^2} \, (0.85 \,
\Delta_1 + 0.91 \, \Delta_2) \approx 0.015 \, \Delta_1 + 0.016 \,
\Delta_2 \approx 0.11 \pm 0.03 \, ps^{-1}~.
\label{dbres}
\end{eqnarray}
When compared with the data on the total decay rate of $\Lambda_b$ this
result predicts about 14\% longer lifetime of $\Xi_b^-$ than that of
$\Lambda_b$.

The singlet matrix elements $x_s$ and $y_s$ (cf. eq.(\ref{xys})) are
related to the shift of the average decay rate of the hyperons in the
triplet:
\beq
{\overline \Gamma}_Q= {1 \over 3} \, \left ( \Gamma(\Lambda_Q) +
\Gamma(\Xi_Q^1)+\Gamma(\Xi_Q^2) \right )~.
\label{avgam}
\eeq
For the charmed baryons the shift of the dominant non-leptonic decay
rate is given by \cite{mv:99-2}
\beq
\delta_{nl}^{(3,0)} {\overline \Gamma}_c= \cos^4 \theta \, {G_F^2 \,
m_c^2 \over 8 \pi} (C_+^2 + C_-^2)\, \kappa^{5/18} \, (x_s-3 \, y_s)~,
\label{d3gc}
\eeq
while for the $b$ baryons the corresponding expression reads as
\beq
\delta^{(3)} {\overline \Gamma}_b= |V_{bc}|^2 \, {G_F^2 \, m_b^2 \over 8
\pi} ({\tilde C}_+ - {\tilde C}_-)^2\, {\tilde \kappa}^{5/18} \, (x_s-3
\, y_s)~.
\label{d3gb}
\eeq

The combination $x_s-3 \, y_s$ of the SU(3) singlet matrix elements
cancels in the ratio of the shifts for $b$ hyperons and the charmed
ones:
\beq
\delta^{(3)} {\overline \Gamma}_b= {|V_{bc}|^2 \over \cos^4} \,
{m_b^2 \over m_c^2} \,
{({\tilde C}_+ - {\tilde C}_-)^2 \over C_+^2 + C_-^2 } \,  \left [
{\alpha_s (m_c) \over \alpha_s(m_b)} \right ]^{5/18} \,
\delta_{nl}^{(3,0)} {\overline \Gamma}_c \approx 0.0025 \,
\delta_{nl}^{(3,0)} {\overline \Gamma}_c~.
\label{rbc}
\eeq
(One can observe, with satisfaction, that the dependence on the
unphysical parameter $\mu$ cancels out, as it should.)
This equation shows that relatively to the charmed baryons the
shift of the decay rates in the $b$ baryon triplet is greatly suppressed
by the ratio $({\tilde C}_+ - {\tilde C}_-)^2 / (C_+^2 + C_-^2)$, which
parametrically is of the second order in $\alpha_s$, and numerically is
only about 0.12.

An estimate of $\delta^{(3)} {\overline \Gamma}_b$ from eq.(\ref{rbc})
in absolute terms depends on evaluating the average shift
$\delta_{nl}^{(3,0)} {\overline \Gamma}_c$ for charmed baryons. The
latter shift can be conservatively bounded from above by the average
total decay rate of those baryons: $\delta_{nl}^{(3,0)} {\overline
\Gamma}_c < {\overline \Gamma_c} = 6.0 \pm 0.7 \, ps^{-1}$, which then
yields, using eq.(\ref{rbc}), an upper bound $\delta^{(3)} {\overline
\Gamma}_b < 0.015 \pm 0.002 \, ps^{-1}$. More realistically, one should
subtract from the total average width ${\overline \Gamma_c}$ the
contribution of the `parton' term, which can be estimated from the decay
rate of $D_0$ with account of the $O(m_c^{-2})$ effects, as
amounting to about $3 \, ps^{-1}$. (One should also take into account
the semileptonic contribution to the total decay rates, which however is
quite small at this level of accuracy). Thus a realistic evaluation of
$\delta^{(3)} {\overline \Gamma}_b$ does not exceed $0.01 \, ps^{-1}$,
which constitutes only about 1\% of the total decay rate of $\Lambda_b$.
Thus the shift of the total decay rate of $\Lambda_b$ due to the effects
of $L_{eff}^{(3)}$ is dominantly associated with the SU(3) non-singlet
difference (\ref{dbres}). The shift of the $\Lambda_b$ decay rate with
respect to the average width ${\overline \Gamma}_b$ due to the
non-singlet operators is one third of the splitting (\ref{dbres}), i.e.
about 5\%. Adding to this the 1\% shift of the average width and another
1\% difference from the meson decays due to the suppression of the
latter by the $m_b^{-2}$ chromomagnetic effects, one concludes that at
the present level of theoretical understanding it looks impossible to
explain a more than 10\% enhancement of the total decay rate of
$\Lambda_b$ relative to $B_d$, where an ample 3\% margin is added for
the uncertainties of higher order terms in OPE as well as for higher
order QCD radiative effects in the discussed corrections. In other
words, the expected pattern of the lifetimes of the $b$ hyperons in the
triplet, relative to $B_d$, is
\beq
\tau(\Xi_b^0) \approx \tau(\Lambda_b) < \tau(B_d) < \tau(\Xi_b^-)~,
\label{pattern}
\eeq
with the ``best" theoretical estimate of the differences to be about 7\%
for each step of the inequality.

For the double strange hyperons $\Omega_c$ and $\Omega_b$ there is
presently no better approach to evaluating the four-quark matrix
elements, than the use of simplistic relations, like (\ref{omegaold})
based on constituent quark model. Such relations imply that the effects
of the strange quark, WS and PI, in the $\Omega_Q$ baryons are
significantly enhanced over the same effects in the cascade hyperons. In
charmed baryons a presence of strange spectator quark enhances the decay
through positive interference with the quark emerging from the $c \to s$
transition in the decay. For $\Omega_c$ this implies a significant
enhancement of the total decay rate \cite{sv:86}, which is in perfect
agreement with the data on the $\Omega_c$ lifetime. Also a similar
enhancement is expected for the semileptonic decay rate of $\Omega_c$.
In $b$ baryons, on the contrary, the interference effect for a spectator
strange quark is negative. Thus the non-leptonic decay rate of
$\Omega_b$ is expected to be suppressed, leaving $\Omega_b$ most
probably the longest-living particle among the $b$ baryons.

\section{Relation between spectator effects in baryons and the decays
$\Xi_Q \to \Lambda_Q \, \pi$}
Rather unexpectedly, the problem of four-quark matrix elements over
heavy hyperons is related to decays of the type $\Xi_Q \to \Lambda_Q \,
\pi$.
The mass difference between the charmed cascade hyperons $\Xi_c$ and
$\Lambda_c$ is about 180 MeV. The expected analogous mass splitting for
the $b$ hyperons should be very close to this number, since in the heavy
quark limit
\beq
M(\Xi_b)-M(\Lambda_b)=M(\Xi_c)-M(\Lambda_c) + O(m_c^{-2}-m_b^{-2}).
\label{massdiff}
\eeq
Therefore in both cases are kinematically possible decays of the type
$\Xi_Q \to \Lambda_Q \, \pi$, in which the heavy quark is not destroyed,
and which are quite similar to decays of ordinary `light' hyperons.
Surprisingly, the rate of these decays for both $\Xi_c$ and $\Xi_b$ is
not insignificantly small, but rather their branching fraction can reach
a level of few per mill for $\Xi_c$ and of one percent or
more for $\Xi_b$ \cite{mv:00}.

The transitions $\Xi_Q \to \Lambda_Q \, \pi$ are induced by two
underlying weak processes: the `spectator' decay of a strange quark,
$s \to u \, \overline u \, d$, which does not involve the heavy quark,
and the `non-spectator' weak scattering (WS)
\beq
s \, c \to c \, d
\label{ws}
\eeq
trough the weak interaction of the $c \to d$ and $s \to c$ currents.
One can also readily see that the WS mechanism contributes only to the
decays $\Xi_c \to \Lambda_c \, \pi$ and is not present in the decays of
the $b$ cascade hyperons. An important starting point in considering
these transitions is that in the heavy quark limit the spin of the heavy
quark completely decouples from the spin of the light component of the
baryon, and that the latter light component in both the initial an the
final baryons forms a $J^P=0^+$ state with quantum numbers of a diquark.
Since the momentum transfer in the considered decays is small in
comparison with the mass of the heavy quark the spin of the amplitudes
with spin flip of the heavy quark, and thus of the baryon, are
suppressed by $m_Q^{-1}$. In terms of the two possible partial waves in
the decay $\Xi_Q \to \Lambda_Q \, \pi$, the $S$ and $P$, this implies
that the $P$ wave is strongly suppressed and the decays are dominated by
the $S$ wave.

According to the well known current algebra
technique, the $S$ wave amplitudes of pion emission can be considered in
the chiral limit at zero four-momentum of the pion, where they are
described by the PCAC reduction formula (pole terms are absent in these
processes):
\beq
\langle \Lambda_Q \, \pi_i (p=0) \,| H_W |\, \Xi_Q \rangle = {\sqrt{2}
\over f_\pi} \, \langle \Lambda_Q \, |\left[Q^5_i, \, H_W \right ] |\,
\Xi_Q \rangle~,
\label{pcac}
\eeq
where $\pi_i$ is the pion triplet in the Cartesian notation, and $Q^5_i$
is the corresponding isotopic triplet of axial charges. The constant
$f_\pi \approx 130 \, MeV$, normalized by the charged pion decay, is
used here, hence the coefficient $\sqrt{2}$ in eq.(\ref{pcac}). The
Hamiltonian $H_W$ in eq.(\ref{pcac}) is the non-leptonic
strangeness-changing hamiltonian:
\begin{eqnarray}
H_W = &&\sqrt{2} \, G_F \, \cos \theta_c \, \sin \theta_c \left \{ \left
( C_+ + C_-
\right ) \, \left [ (\overline u_L \, \gamma_\mu \, s_L)\, (\overline
d_L \, \gamma_\mu \, u_L) - (\overline c_L \, \gamma_\mu \, s_L)\,
(\overline d_L \, \gamma_\mu \, c_L) \right ] + \right . \nonumber \\
&& \left. \left ( C_+ - C_- \right ) \, \left [ (\overline d_L \,
\gamma_\mu \, s_L)\, (\overline u_L \, \gamma_\mu \, u_L) - (\overline
d_L \, \gamma_\mu \, s_L)\, (\overline c_L \, \gamma_\mu \, c_L) \right
] \right \}~.
\label{hw}
\end{eqnarray}
In this formula the weak Hamiltonian is assumed to be normalized (in
LLO) at $\mu = m_c$.
The terms in the Hamiltonian (\ref{hw}) without the charmed quark fields
describe the `spectator' nonleptonic decay of the strange quark, while
those with the $c$ quark correspond to the WS process (\ref{ws}).

It is straightforward to see from eq.(\ref{pcac}) that in the PCAC limit
the discussed decays should obey the $\Delta I =1/2$ rule. Indeed, the
commutator of the weak Hamiltonian with the axial charges transforms
under the isotopic SU(2) in the same way as the Hamiltonian itself. In
other words, the $\Delta I=1/2$ part of $H_W$ after the commutation
gives an $\Delta I=1/2$ operator, while the $\Delta I = 3/2$ part after
the commutation gives an $\Delta I = 3/2$ operator. The latter operator
however cannot have a non vanishing matrix element between an isotopic
singlet, $\Lambda_Q$, and an isotopic doublet, $\Xi_Q$. Thus the $\Delta
I=3/2$ part of $H_W$ gives no contribution to the $S$ wave amplitudes in
the PCAC limit.

Once the isotopic properties of the decay amplitudes are fixed, one can
concentrate on specific charge decay channels, e.g. $\Xi_b^- \to
\Lambda_b \, \pi^-$ and $\Xi_c^0 \to \Lambda_c \, \pi^-$. An application
of the PCAC relation (\ref{pcac}) with the Hamiltonian from
eq.(\ref{hw}) to these decays, gives the expressions for the amplitudes
at $p=0$ in terms of baryonic matrix elements of four-quark operators:
\begin{eqnarray}
&&\langle \Lambda_b \, \pi^- (p=0) \,| H_W |\, \Xi_b^- \rangle =
\nonumber \\
&&{\sqrt{2} \over f_\pi} \, G_F \, \cos \theta_c \, \sin \theta_c \,
\langle \Lambda_b
\,| \left ( C_+ + C_- \right ) \, \left [ (\overline u_L \, \gamma_\mu
\, s_L)\, (\overline d_L \, \gamma_\mu \, d_L) - (\overline u_L \,
\gamma_\mu \, s_L)\, (\overline u_L \, \gamma_\mu \, u_L) \right ] +
\nonumber \\
&& \left ( C_+ - C_- \right ) \, \left [ (\overline d_L \, \gamma_\mu \,
s_L)\, (\overline u_L \, \gamma_\mu \, d_L) - (\overline u_L \,
\gamma_\mu \, s_L)\, (\overline u_L \, \gamma_\mu \, u_L) \right] | \,
\Xi_b^- \rangle = \nonumber \\
&&{\sqrt{2} \over f_\pi} \, G_F \, \cos \theta_c \, \sin \theta_c \,
\langle \Lambda_b
\,| C_- \, \left [ (\overline u_L \, \gamma_\mu \, s_L)\, (\overline d_L
\, \gamma_\mu \, d_L) - (\overline d_L \, \gamma_\mu \, s_L)\,
(\overline u_L \, \gamma_\mu \, d_L) \right ] - \nonumber \\
&& {C_+ \over 3} \left [ (\overline u_L \, \gamma_\mu \, s_L)\,
(\overline d_L \, \gamma_\mu \, d_L) + (\overline d_L \, \gamma_\mu \,
s_L)\, (\overline u_L \, \gamma_\mu \, d_L) +2 \, (\overline u_L \,
\gamma_\mu \, s_L)\, (\overline u_L \, \gamma_\mu \, u_L) \right ] | \,
\Xi_b^- \rangle~,
\label{xib}
\end{eqnarray}
where in the last transition the operator structure with $\Delta I =3/2$
giving a vanishing contribution is removed and only the structures with
explicitly $\Delta I =1/2$ are retained, and
\begin{eqnarray}
&&\langle \Lambda_c \, \pi^- (p=0) \,| H_W |\, \Xi_c^0 \rangle = \langle
\Lambda_b \, \pi^- (p=0) \,| H_W |\, \Xi_b^- \rangle +  \nonumber \\
&&{\sqrt{2} \over f_\pi} \, G_F \, \cos \theta_c \, \sin \theta_c \,
\langle \Lambda_c \,
| \left ( C_+ + C_- \right ) \, (\overline c_L \, \gamma_\mu \, s_L)\,
(\overline u_L \, \gamma_\mu \, c_L) + \nonumber \\
&&\left ( C_+ - C_- \right ) \, (\overline u_L \, \gamma_\mu \, s_L)\,
(\overline c_L \, \gamma_\mu \, c_L) |\, \Xi_c^0 \rangle~.
\label{xic}
\end{eqnarray}
In the latter formula the first term on the r.h.s. expresses the fact
that in the heavy quark limit the `spectator' amplitudes do not depend
on the flavor or the mass of the heavy quark. The rest of the expression
(\ref{xic}) describes the
`non-spectator' contribution to the amplitude of the charmed hyperon
decay. Using the flavor SU(3) symmetry the latter contribution can be
related
to the non-singlet matrix elements (\ref{defxy}) (normalized at
$\mu=m_c$) as
\begin{eqnarray}
\Delta A \equiv \langle \Lambda_c \, \pi^- (p=0) \,| H_W |\, \Xi_c^0
\rangle - \langle \Lambda_b \, \pi^- (p=0) \,| H_W |\, \Xi_b^- \rangle =
\nonumber \\
{G_F \cos \theta_c \, \sin \theta_c \over 2 \, \sqrt{2} \, f_\pi} \,
\left[ \left ( C_- -
C_+ \right ) \, x - \left ( C_+ + C_- \right ) \, y \right ]~.
\label{das}
\end{eqnarray}
Furthermore, with the help of the equations (\ref{d1}) and (\ref{d2})
relating the matrix elements $x$ and $y$ to the differences of the total
decay widths within the triplet of charmed hyperons, one can eliminate
$x$ and $y$ in favor of the measured width differences. The resulting
expression has the form
\begin{eqnarray}
\Delta A \approx - {\sqrt{2} \, \pi \, \cos \theta_c \, \sin \theta_c
\over G_F \,
m_c^2 \, f_\pi} \, \left [ 0.45 \, \left (
\Gamma(\Xi_c^0)-\Gamma(\Lambda_c) \right ) + 0.04 \left (
\Gamma(\Lambda_c)-\Gamma(\Xi_c^+) \right ) \right ] = \nonumber \\
-10^{-7} \left [
0.97 \left ( \Gamma(\Xi_c^0)-\Gamma(\Lambda_c) \right ) + 0.09 \left (
\Gamma(\Lambda_c)-\Gamma(\Xi_c^+) \right ) \right ] \, \left ( {1.4 \,
GeV \over m_c} \right )^2 \, ps ~,
\label{dasm}
\end{eqnarray}
where, clearly, in the latter form the widths are assumed to be
expressed in $ps^{-1}$, and $m_c=1.4 \, GeV$ is used as a `reference'
value for the charmed quark mass. It is seen from eq.(\ref{dasm}) that
the evaluation of the difference of the amplitudes within the discussed
approach is mostly sensitive to the difference of the decay rates of
$\Xi_c^0$ and $\Lambda_c$, with only very little sensitivity to the
total decay width of $\Xi_c^+$. Using the current data  the difference
$\Delta A$ is estimated as
\beq
\Delta A= -(5.4 \pm 2) \times 10^{-7}~,
\label{dasn}
\eeq
with the uncertainty being dominated by the experimental error in the
lifetime of $\Xi_c^0$. An  amplitude $A$ of the magnitude,
given by the central value in eq.(\ref{dasn}) would produce a decay rate
$\Gamma (\Xi_Q \to \Lambda_Q \, \pi) = |A|^2 \, p_\pi/(2 \pi) \approx
0.9 \times 10^{10} \, s^{-1}$, which result can also be written in a
form of triangle inequality
\beq
\sqrt{\Gamma(\Xi_b^- \to \Lambda_b \, \pi^-)} + \sqrt{\Gamma(\Xi_c^0 \to
\Lambda_c \, \pi^-)} \ge \sqrt{0.9 \times 10^{10} \, s^{-1}}~.
\label{tri}
\eeq

Although at present it is not possible to evaluate in a reasonably model
independent way the matrix element in eq.(\ref{xib}) for the `spectator'
decay amplitude, the inequality (\ref{tri}) shows that at least some of
the discussed pion transitions should go at the level of $0.01 \,
ps^{-1}$, similar to the rates of analogous decays of `light' hyperons.

\section{Summary}

We summarize here specific predictions for the inclusive decay rates,
which can be argued with a certain degree of theoretical reliability,
and which can be possibly experimentally tested in the nearest future.

$B$ mesons:
$$ \tau(B_d)/\tau(B_s)=1 \pm 0.01~.$$

 Charmed hyperons:
$$\Gamma_{sl}(\Xi_c) = (2 - 3) \, \Gamma_{sl}(\Lambda_c)\, ~~~~
\Gamma_{sl}(\Omega_c) > \Gamma_{sl}(\Xi_c)~,$$
$$\Gamma^{nl}_{\Delta S =
-1}(\Xi_c^+) \approx \Gamma^{nl}_{\Delta S = -1}(\Lambda_c) ~,$$
$$\Gamma^{nl}_{\Delta S =
-1}(\Xi_c^0)-\Gamma^{nl}_{\Delta S = -1}(\Lambda_c) \approx 0.55 \pm
0.22 \, ps^{-1}~.$$

$b$ hyperons:
$$\tau(\Xi_b^0) \approx \tau(\Lambda_b) < \tau(B_d) <
\tau(\Xi_b^-)<\tau(\Omega_b)~,$$
$$\Gamma(\Lambda_b) - \Gamma(\Xi_b^-) \approx 0.11 \pm 0.03 \,
ps^{-1}~,$$
$$0.9 < {\tau(\Lambda_b) \over \tau(B_d)} < 1~.$$

Strangeness decays $\Xi_Q \to \Lambda_Q \, \pi$: \\
The $\Delta I =1/2$ rule should hold in these decays, so that
$\Gamma(\Xi_Q^{(d)} \to \Lambda_Q \pi^-)  = 2 \, \Gamma(\Xi_Q^{(u)} \to
\Lambda_Q \pi^0)$. The rates are constrained by the triangle inequality
(\ref{tri}).\\[0.2in]

\noindent
{\large \bf Acknowledgement}

\noindent
This work is supported in part by DOE under the grant number
DE-FG02-94ER40823.

\end{document}